\documentclass{webofc}

\usepackage[varg]{txfonts}   
\usepackage{hyperref}
\usepackage{url}

\newcommand{\NASixtyOne}{NA61\slash SHINE\xspace}
\newcommand{\byc}{\kern-0.1em/\kern-0.1em c}
\newcommand{\GeV}{\ensuremath{\mbox{Ge\kern-0.1em V}}\xspace}
\newcommand{\MeV}{\ensuremath{\mbox{Me\kern-0.1em V}}\xspace}
\newcommand{\GeVc}{\ensuremath{\mbox{Ge\kern-0.1em V}\byc}\xspace}
\newcommand{\AGeVc}{\ensuremath{A\,\mbox{Ge\kern-0.1em V}\byc}\xspace}
\newcommand{\MeVc}{\ensuremath{\mbox{Me\kern-0.1em V}\byc}\xspace}

\newcommand{\mt}{\ensuremath{m_{\textrm T}}\xspace}

\newcommand{\Ks}{\ensuremath{K^0_\mathrm{S}}\xspace}
\newcommand{\Kl}{\ensuremath{K^0_\mathrm{L}}\xspace}
\newcommand{\pp}{\mbox{\textit{p+p}}\xspace}

\newcommand{\snn}{\ensuremath{\sqrt{s_{\mathrm{NN}}}}\xspace}
\usepackage{comment}

\hypersetup{colorlinks=true,citecolor=blue,urlcolor=blue,linkcolor=blue}
\begin{document}
\title{News from NA61/SHINE}
%
%

\author{\firstname{Katarzyna} \lastname{Grebieszkow}\inst{1}\fnsep\thanks{\email{katarzyna.grebieszkow@pw.edu.pl}} for the NA61/SHINE Collaboration  
}

\institute{Warsaw University of Technology, Faculty of Physics, Koszykowa 75, 00-662 Warsaw, Poland}

\abstract{
The main goal of the \NASixtyOne strong interaction program is to search for the critical point in the phase diagram of strongly interacting matter and to investigate phenomena related to the onset of deconfinement. In recent years, the program has expanded to include the study of open charm, aiming to understand the mechanisms of its production in heavy-ion collisions. This article presents a selection of recent results from the \NASixtyOne strong interaction program, including findings on particle spectra and yields, as well as fluctuations and correlations. Plans for the near future are also discussed.
}

\maketitle
%

\section{Introduction}
\label{sec:intro}

\NASixtyOne is a multipurpose fixed-target experiment located at the CERN Super Proton Synchrotron (SPS). It has a rich experimental program covering strong interaction, neutrino, and cosmic-ray physics. Within the neutrino program, \NASixtyOne provides hadron production measurements essential for neutrino experiments at J-PARC and Fermilab. Under the cosmic-ray program, the experiment delivers data for interpreting cosmic-ray showers, supporting projects such as Pierre Auger, KASCADE, and satellite-based experiments. For the strong interaction (SI) program, \NASixtyOne studies the properties of the onset of deconfinement (in principle, the possibility of a hadron gas $\leftrightarrow$ quark-gluon plasma transition is studied for light and intermediate-mass systems), searches for the critical point in the phase diagram of strongly interacting matter, and contributes to understanding the mechanisms of open charm production in heavy-ion collisions. The SI program is based on systematic beam momentum scans (13$A$--150$A$/158\AGeVc; collision center-of-mass energy per nucleon pair \snn = 5.1--16.8/17.3~\GeV) involving different mass nuclei, ranging from \pp to Pb+Pb.

\section{Properties of the onset of deconfinement}
\label{sec:OD}

\begin{figure}[h]
\centering
\includegraphics[width=4.25cm,clip]{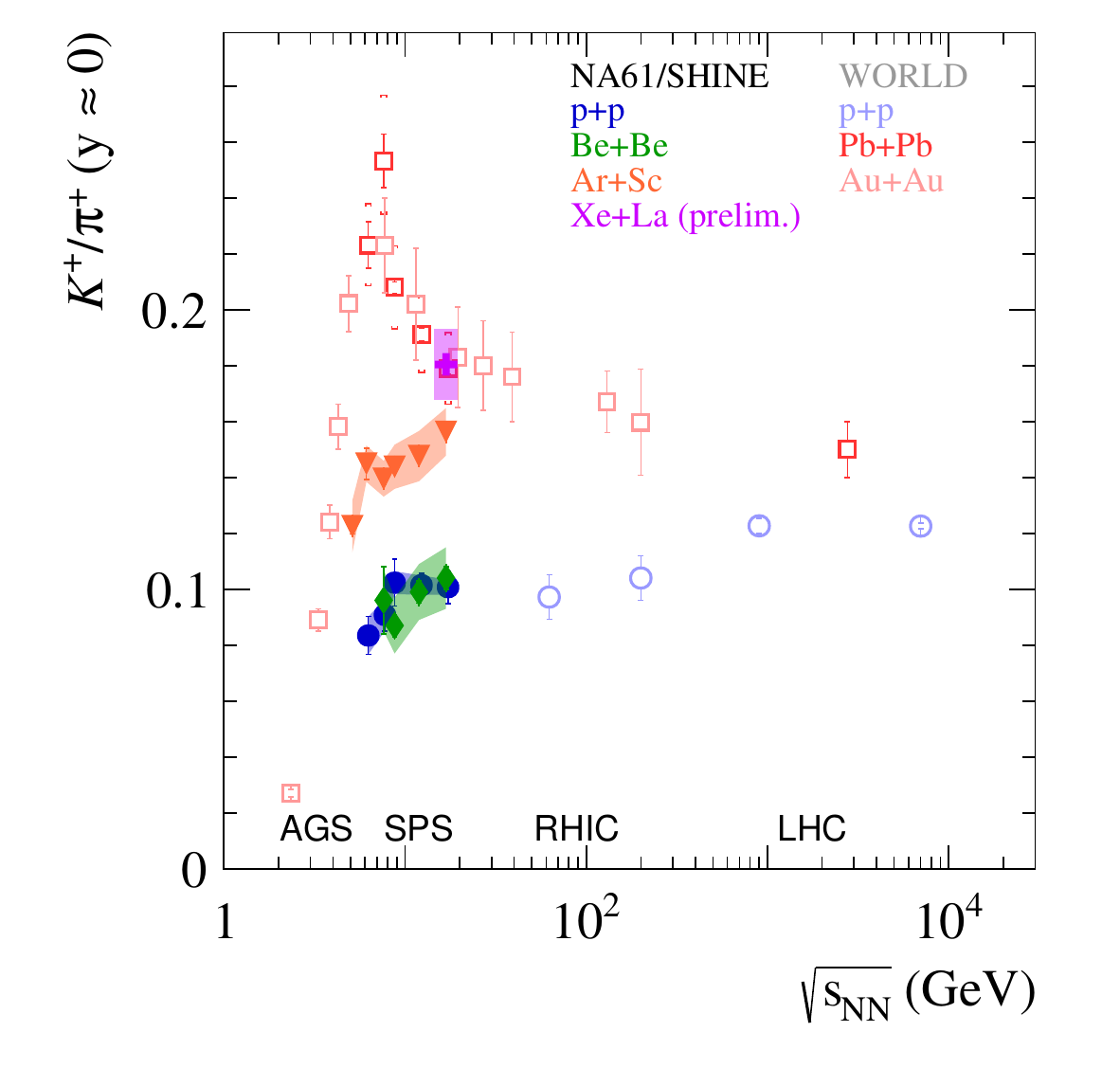}
\includegraphics[width=4.25cm,clip]{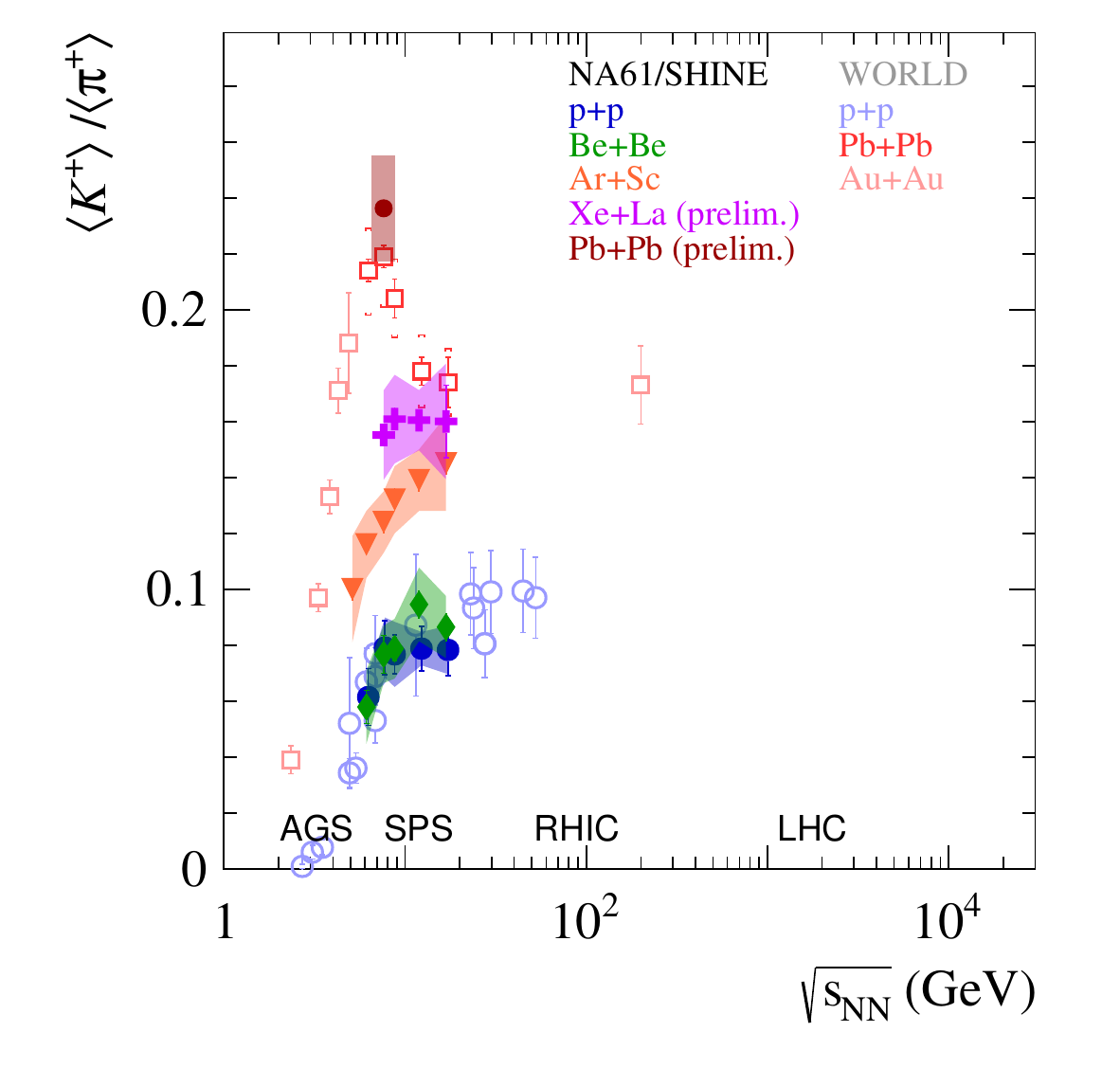} 
\includegraphics[width=4.25cm,clip]{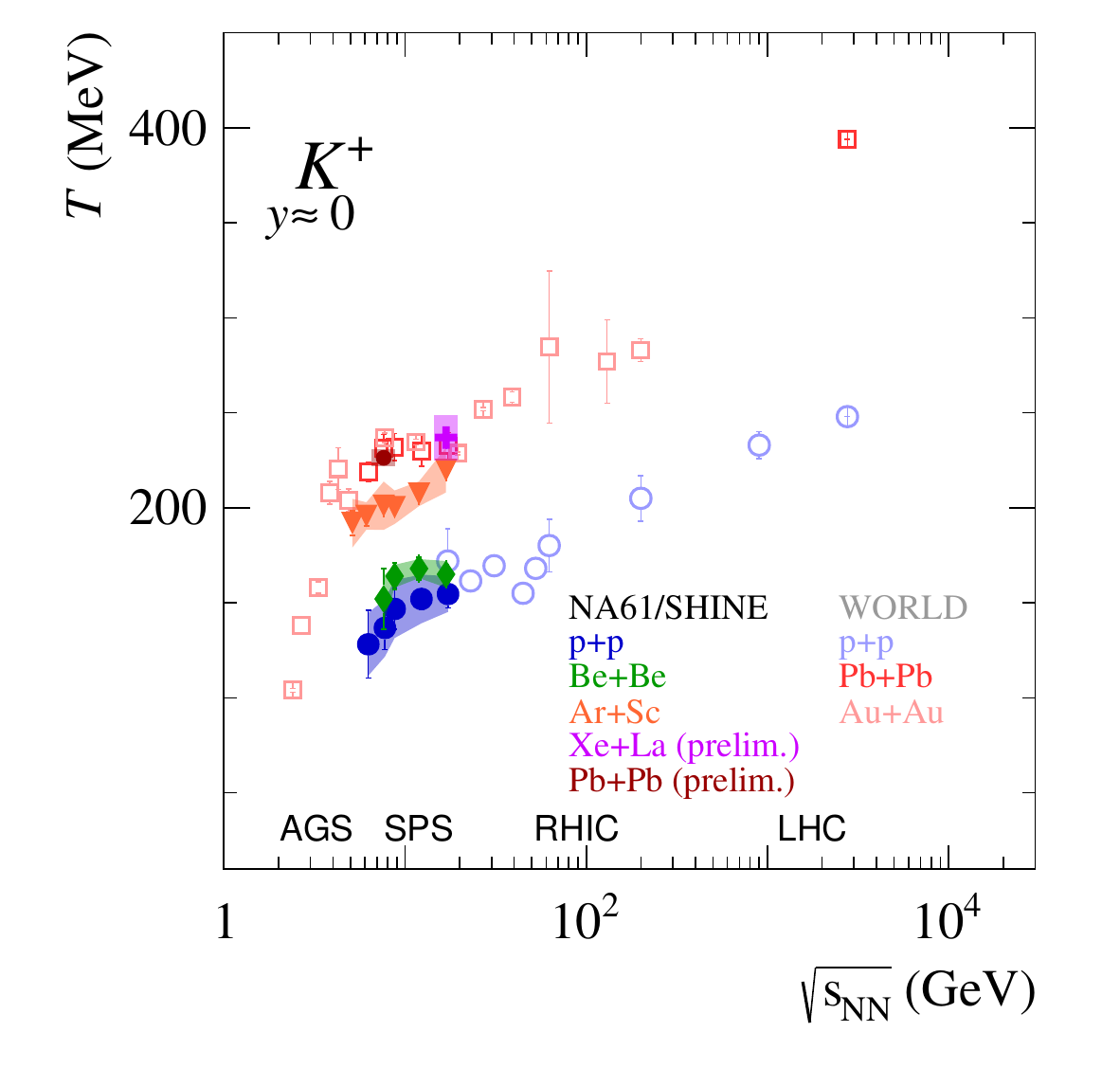}
\vspace{-0.35cm}
\caption{Left and middle: horn plots -- $K^{+}/\pi^{+}$ ratio versus energy at mid-rapidity (left) and in $4\pi$ (middle). Right: example of step plot -- inverse slope parameter of transverse mass/transverse momentum spectrum of $K^{+}$. See Ref.~\cite{NA61SHINE:2023epu} for 0--10\% Ar+Sc results and references to \NASixtyOne published \pp and 0--20\% Be+Be as well as world data. \NASixtyOne Xe+La (0--20\% central for 16.8~\GeV and 0--10\% central for lower energies) and 0--7.2\% Pb+Pb results are preliminary. For Xe+La, $T$ and $K^{+}/\pi^{+}$ at mid-rapidity ($y \approx 0$) was in fact obtained for $0.4 < y < 0.6$. For \NASixtyOne Pb+Pb, $T$ was obtained for $0.8 < y < 1.0$. For \NASixtyOne points, vertical bars denote statistical uncertainties, and
color bands – systematic ones. }
\label{fig:horn_step}
\end{figure}

The spectra and yields analyzed by \NASixtyOne enable the study of the properties of the onset of deconfinement (OD) -- threshold for quark-gluon plasma (QGP) creation -- by examining whether the characteristic structures known as the \textit{kink}, \textit{horn}, and \textit{step} \cite{Gazdzicki:1998vd} also appear in collisions involving small and intermediate-mass nuclei. In the case of Pb+Pb interactions, NA49 observed a sharp peak (horn) in the $K^{+}/\pi^{+}$ ratio (related to strangeness to entropy), interpreted as due to OD~\cite{NA49:2007stj}. Furthermore, a plateau (step) in the inverse slope parameter ($T$) of the exponential transverse mass (\mt) distribution was reported, which may reflect constant temperature and pressure in a mixed phase. The status of the horn and step plots, including \NASixtyOne results for \pp, 0--20\% central Be+Be, 0--10\% central Ar+Sc, 0--10(20)\% central Xe+La, and 0--7.2\% central Pb+Pb, is presented in Fig.~\ref{fig:horn_step}.
New \NASixtyOne Pb+Pb data at \snn = 7.6~\GeV confirm the NA49 results for $T$ and $K^{+}/\pi^{+}$ ratio. A step-like structure in $T$ is visible for all systems. The \pp results are close to those for Be+Be, while Ar+Sc values are higher (though no horn structure is observed). The new Xe+La results (the two lowest energies will be added soon) at the highest energy approach those for Pb+Pb.

\begin{figure}[h]
\centering
\includegraphics[width=4.25cm,clip]{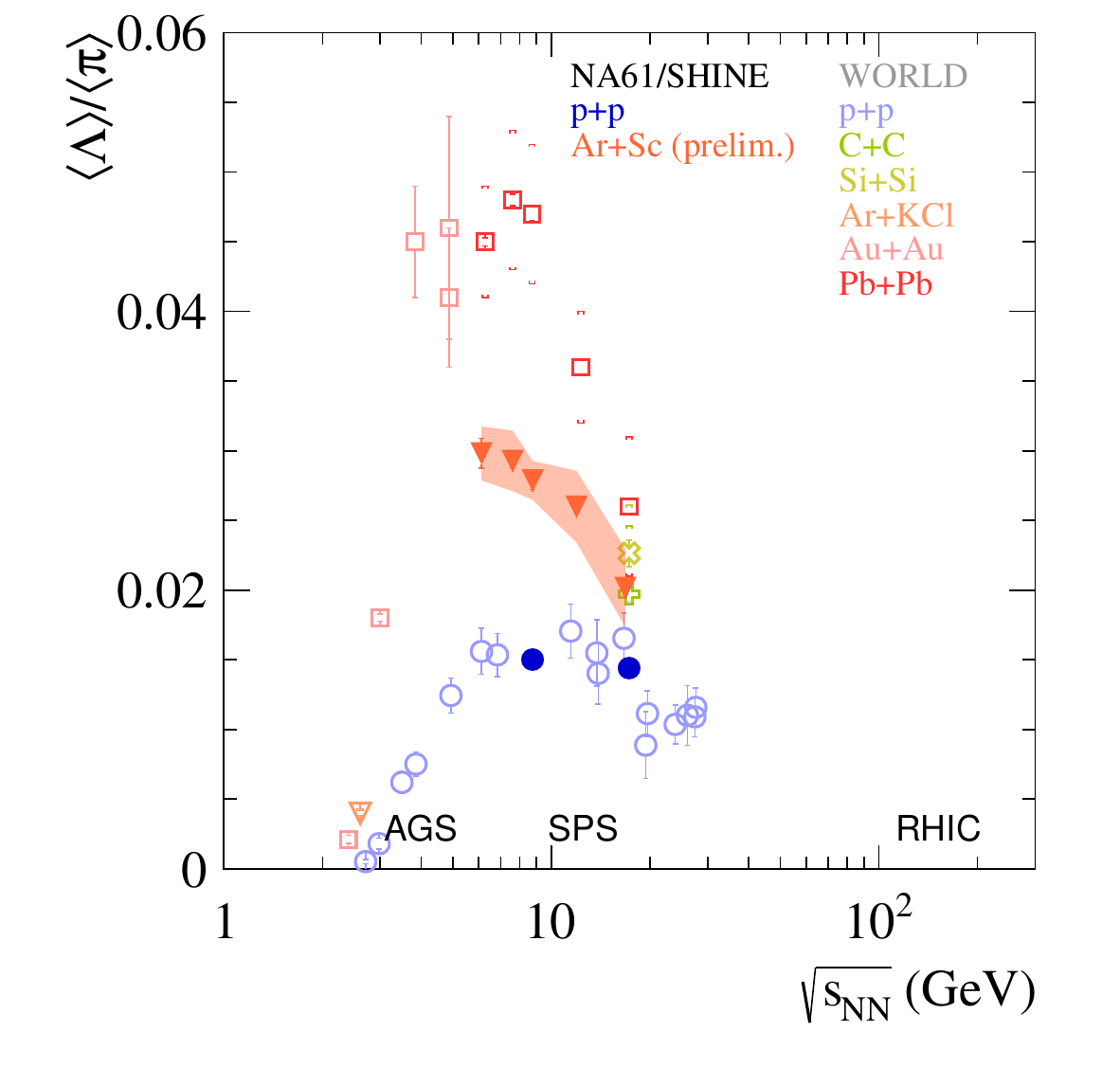}
\includegraphics[width=4.25cm,clip]{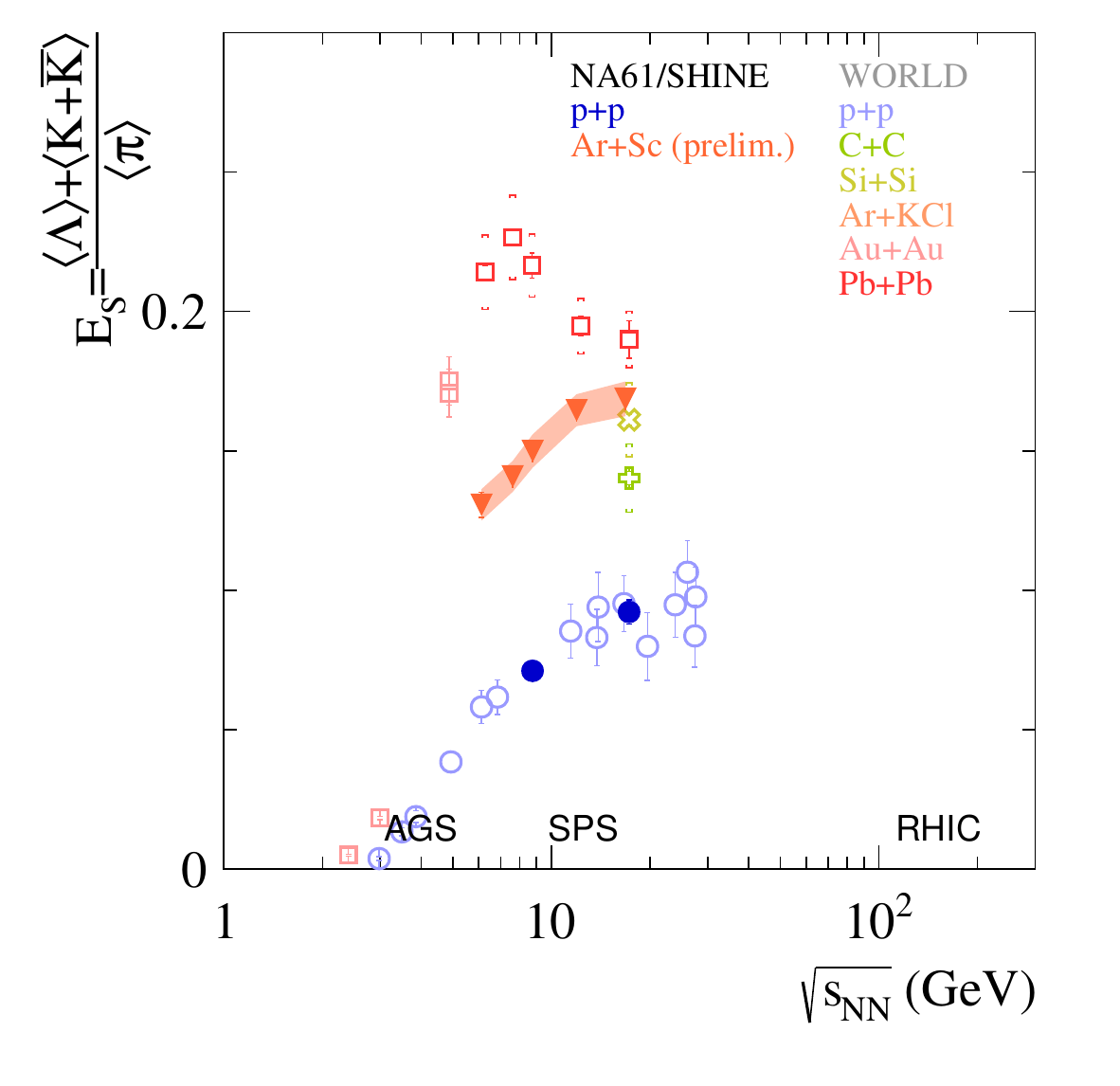}
\vspace{-0.35cm}
\caption{Energy dependence of $\langle \Lambda \rangle / \langle \pi \rangle$ mean multiplicity ratio (left) and $E_\mathrm{S}$ factor (right). \NASixtyOne \pp data are preliminary (8.8~\GeV) and taken from Ref.~\cite{NA61SHINE:2015haq} (17.3~\GeV). \NASixtyOne 0--10\% central Ar+Sc points are preliminary. See Ref.~\cite{QM2025_YB} for references to world data. Assumptions used in the plots: $\langle \pi \rangle = 1.5 \cdot (\langle \pi^+ \rangle + \langle \pi^- \rangle)$; 
$\langle K + \overline{K} \rangle_{A+A} = 2 \cdot (\langle K^+ \rangle + \langle K^- \rangle)$; 
$\langle K + \overline{K} \rangle_{p+p} = 4 \cdot \langle \Ks \rangle$.
For \NASixtyOne points, vertical bars denote statistical uncertainties, and color bands – systematic ones. }
\label{fig:Lambda}
\end{figure}

Recently, \NASixtyOne also obtained preliminary results for $\Lambda$ hyperon production in the 10\% most central Ar+Sc collisions at five SPS energies. A plateau was observed both in the mean multiplicities $\langle \Lambda \rangle$ and in the mid-rapidity yields across the SPS energy range~\cite{QM2025_YB}. Figure~\ref{fig:Lambda} shows the $\langle \Lambda \rangle / \langle \pi \rangle$ ratio as well as the $E_\mathrm{S}$ factor, which similarly to $K^{+}/\pi^{+}$ is approximately proportional to the strangeness to entropy ratio~\cite{Gazdzicki:1998vd} (at SPS energies $\overline{\Lambda}/\Lambda$ ratio is typically smaller than 0.15). As with the $K^{+}/\pi^{+}$ ratio, no horn-like structure is observed in the Ar+Sc $E_\mathrm{S}$ dependence within the SPS energy range. However, the trend of $\langle \Lambda \rangle / \langle \pi \rangle$ in Ar+Sc at SPS is similar to the trend observed in heavy Pb+Pb collisions. The forthcoming data point at the lowest energy (\snn = 5.1~\GeV) will clarify whether a maximum appears in the $\langle \Lambda \rangle / \langle \pi \rangle$ ratio for Ar+Sc collisions.

\section{Search for the critical point}
\label{sec:CP}

Fluctuations and correlations are studied primarily for two reasons. First, they may signal the onset of deconfinement, as the Equation of State changes rapidly near the phase transition, affecting the energy dependence of fluctuations. Second, they can help to locate the critical point (CP) of strongly interacting matter, in analogy to critical opalescence -- enhanced fluctuations near the CP in a liquid-gas transition~\cite{Stephanov:1999zu}. In QCD matter, a maximum of the CP signal is expected when freeze-out occurs close to the CP.
The \NASixtyOne experiment studied, among others, the energy dependence of transverse momentum and multiplicity fluctuations in \pp, Be+Be, and Ar+Sc collisions \cite{Grebieszkow:2017gqx}; the intermittency of protons in Ar+Sc at \snn = 5.1--16.8~\GeV \cite{NA61SHINE:2023gez, SHINE:2024xtq} and in Pb+Pb at \snn = 5.1 and 7.6~\GeV~\cite{Adhikary:2022sdh}; and the intermittency of negatively charged hadrons in Xe+La at \snn = 16.8~\GeV~\cite{ReynaOrtiz:2024hul, QM2025_KG} and in Pb+Pb at \snn = 7.6~\GeV~\cite{Adhikary:2023rfj}. These results did not reveal structures expected for the critical point. In this article, new results from femtoscopic analyses and studies of higher-order moments of net-electric charge distributions are presented. These observables may serve as valuable tools in the search for the critical point.

\begin{figure}[h]
\sidecaption
\centering
\includegraphics[width=6.25cm,clip]{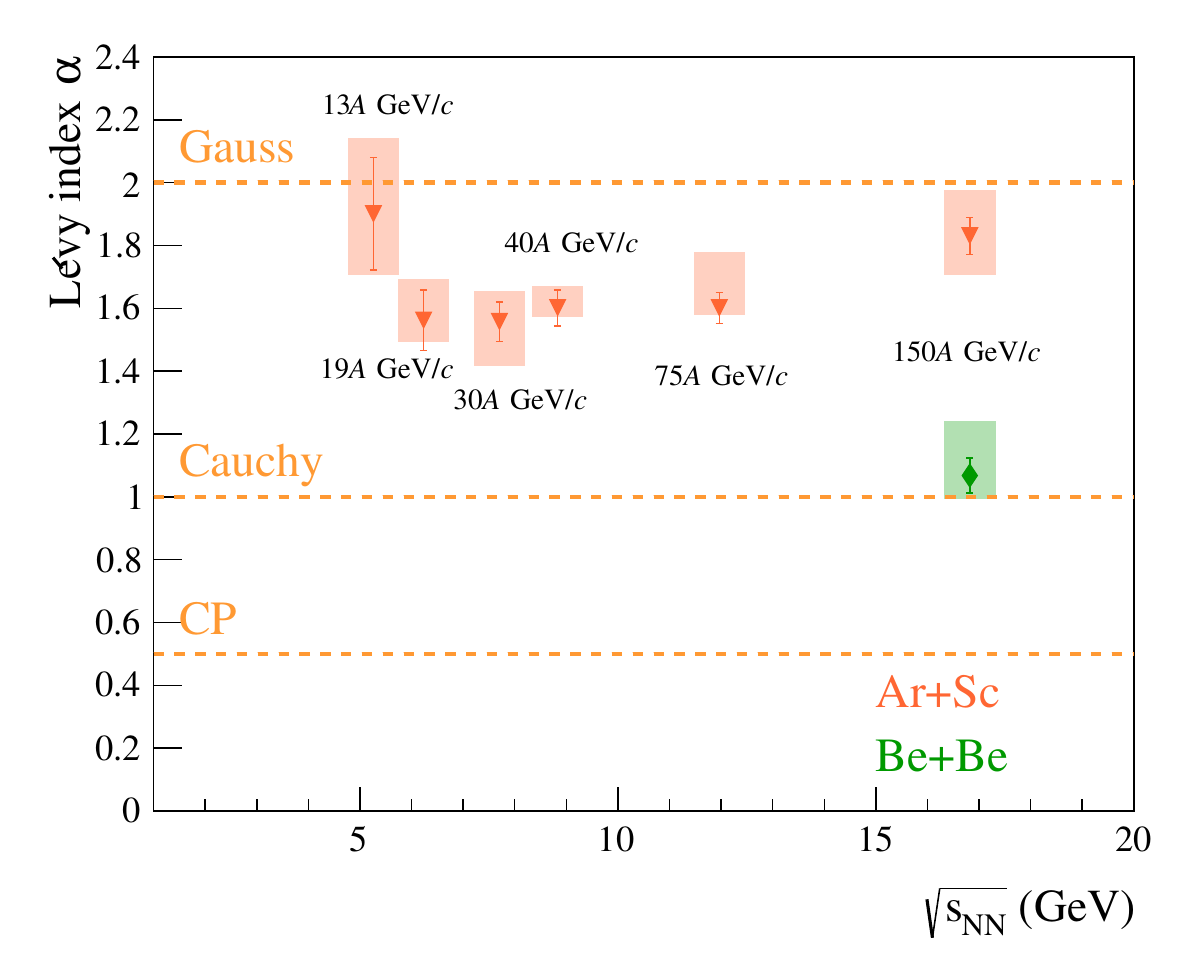}
\caption{L\'evy stability parameter $\alpha$ (constant fit to all \mt values) versus energy in 0--20\% central Be+Be collisions at \snn = 16.8~\GeV (Ref.~\cite{NA61SHINE:2023qzr} with later estimate of systematic uncertainty) as well as in 0--10\% central Ar+Sc collisions at 5.1--16.8~\GeV (\NASixtyOne preliminary). Vertical bars denote statistical uncertainties, and color boxes – systematic ones. }
\label{fig:HBT}
\end{figure}

The femtoscopy analysis of identical (like-sign) charged pions was performed utilizing a L\'evy-type source with the Bose-Einstein part of the two-particle correlation function expressed as $C_2(q)=1+\lambda \cdot e^{-|qR|^\alpha}$, where $q$ is the momentum difference between the particles, $\lambda$ represents the correlation strength, $R$ corresponds to the femtoscopic scale of the system (homogeneity length), and $\alpha$ is the L\'evy stability parameter. The value of $\alpha$ reflects the shape of the emission source: $\alpha$ = 2 corresponds to a Gaussian distribution, $\alpha$ = 1 to a Cauchy distribution, and $\alpha$ $\leq$ 0.5 is expected for the critical point~\cite{PhysRevB.52.6659, NA61SHINE:2023qzr}. 
\NASixtyOne has performed a full energy scan with 0–10\% central Ar+Sc collisions. A constant fit obtained from the transverse mass dependence of $\alpha$ values shows a hint of an interesting minimum at intermediate SPS energies (Fig.~\ref{fig:HBT}). However, the $\alpha$ values remain far from those expected near the critical point, indicating no direct evidence of critical behavior in the studied system. The Be+Be point at the top energy is also far from CP predictions.

\begin{figure}[h]
\centering
\includegraphics[width=4.25cm,clip]{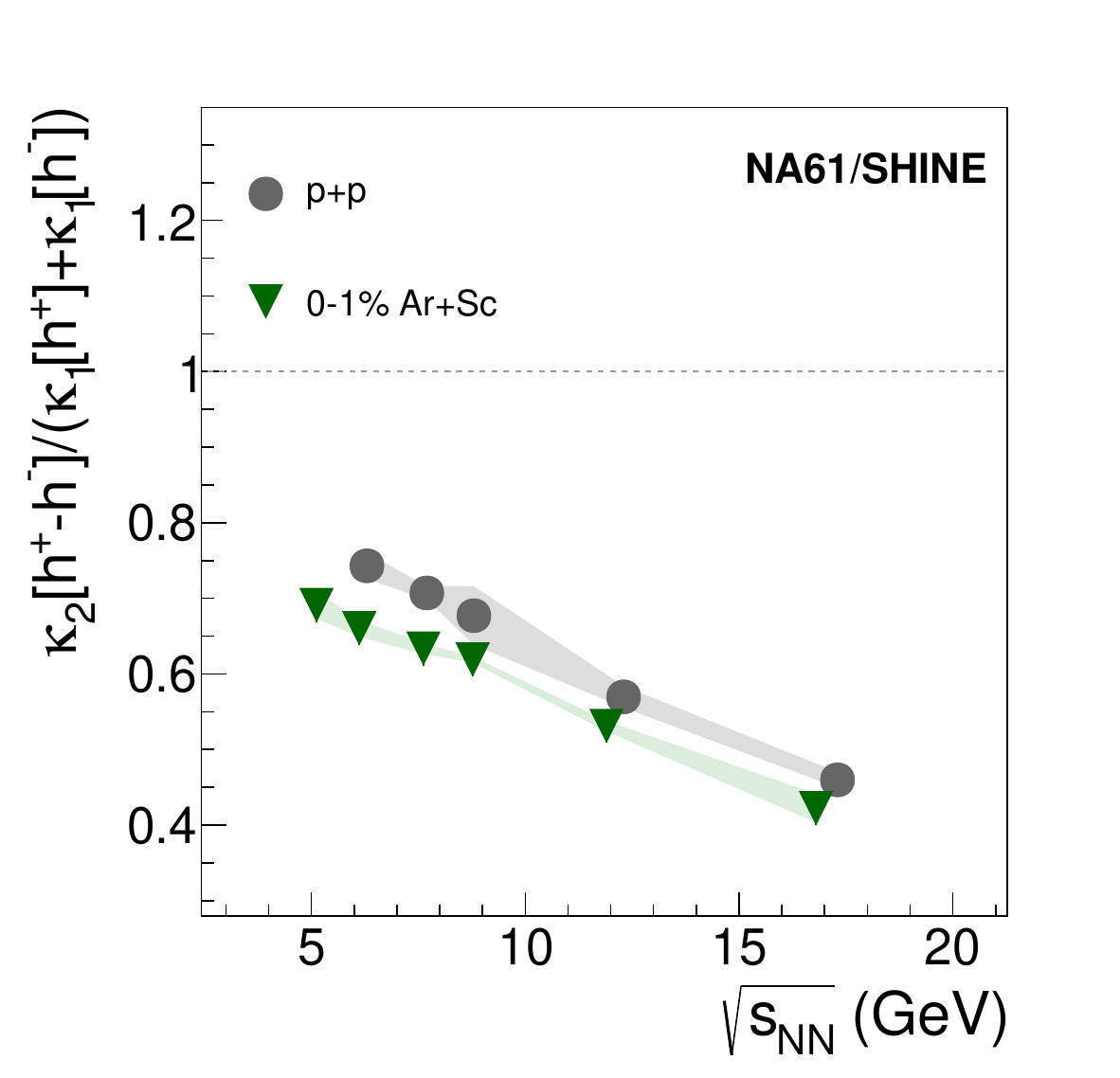}
\includegraphics[width=4.25cm,clip]{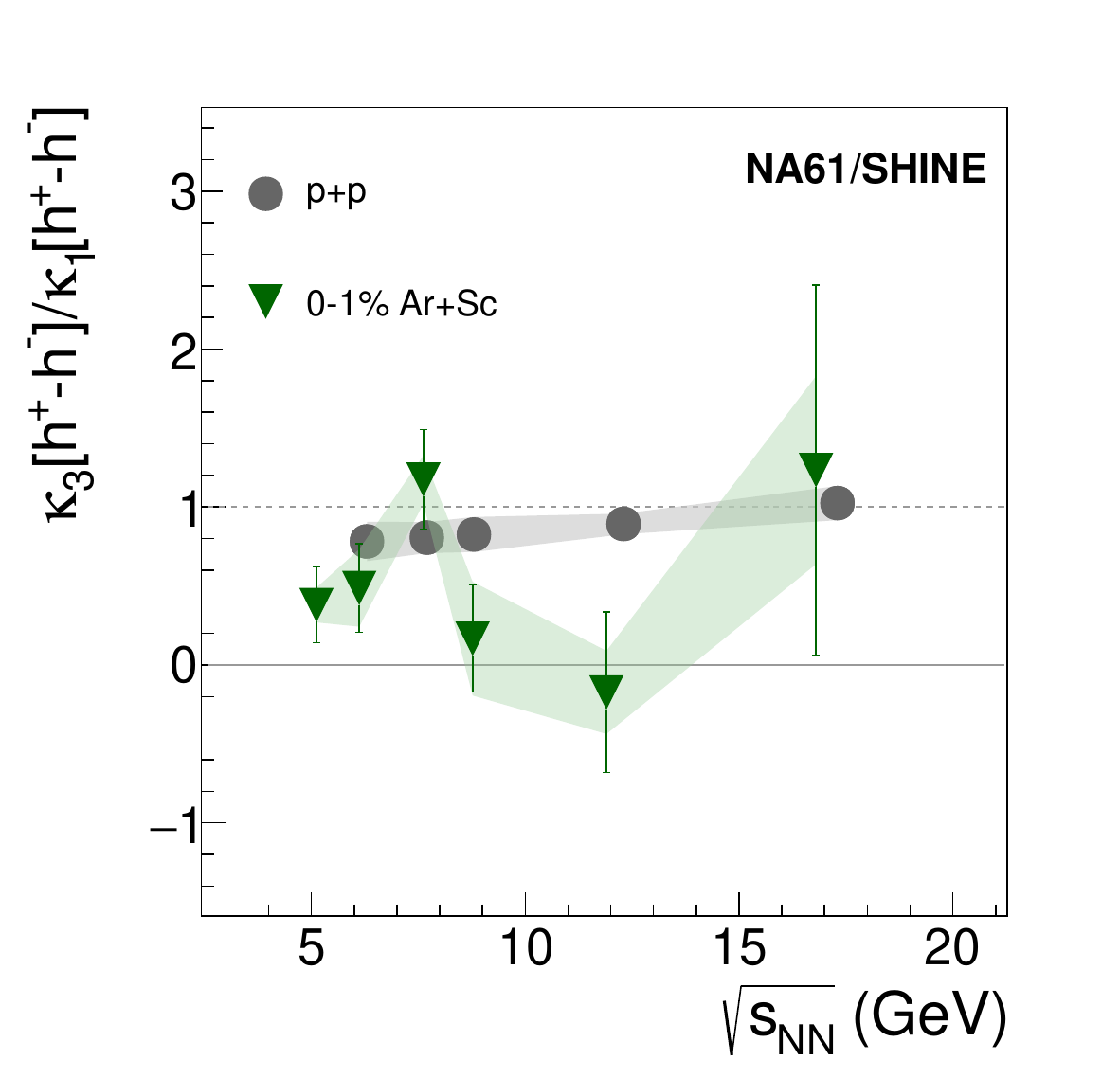}
\includegraphics[width=4.25cm,clip]{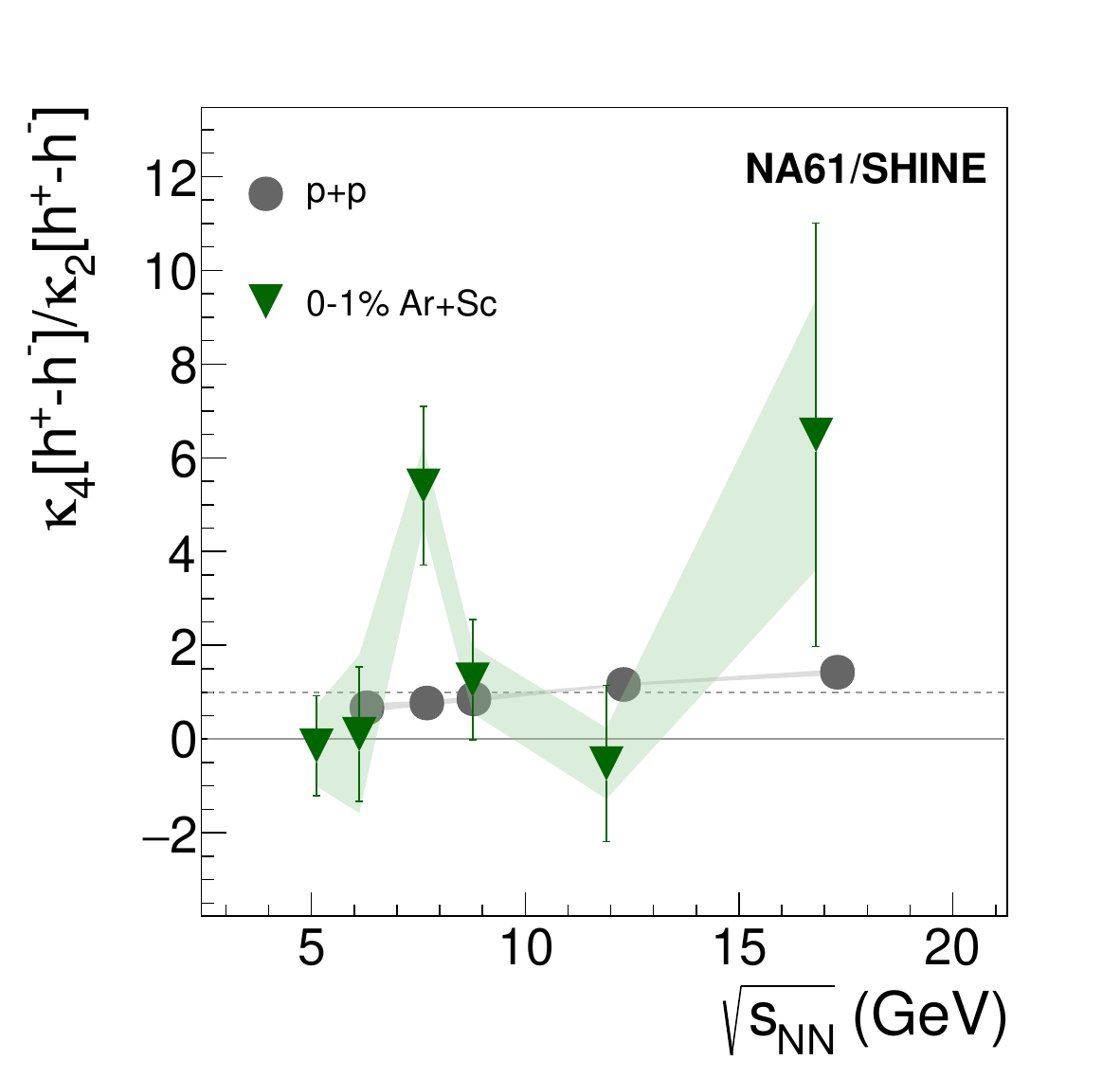}
\vspace{-0.3cm}
\caption{Energy dependence of intensive quantities of net-electric charge distribution in Ar+Sc collisions~\cite{NA61SHINE:2025whi} compared to published \pp results~\cite{SHINE:2023ejm}. 
Vertical bars denote statistical uncertainties, and color bands -- systematic ones. }
\label{fig:results_net}
\end{figure}

\NASixtyOne has recently measured higher-order moments of multiplicity and net-electric charge distributions of hadrons produced in the 1\% most central Ar+Sc collisions at \snn = 5.1--16.8~\GeV~\cite{NA61SHINE:2025whi}. 
Since pure moments are sensitive to the system size, cumulant ($\kappa_n$) ratios are employed instead to probe physical phenomena, particularly the CP signal~\cite{Stephanov:2008qz}. In case of net-electric charge ([$h^{+}-h^{-}$]) we studied $\kappa_{2}[h^{+}-h^{-}]/(\kappa_{1}[h^{+}]+\kappa_{1}[h^{-}])$, $\kappa_{3}/\kappa_{1}[h^{+}-h^{-}]$, and $\kappa_{4}/\kappa_{2}[h^{+}-h^{-}]$. Those measures are intensive and have two reference values: 0 for no fluctuations and 1 for the Skellam distribution. The results are shown in Fig.~\ref{fig:results_net} and compared to \NASixtyOne published \pp points~\cite{SHINE:2023ejm}. A hint of non-monotonic behavior is observed for the ratios $\kappa_{3}/\kappa_{1}$ and $\kappa_{4}/\kappa_{2}$ in Ar+Sc collisions. However, substantial statistical uncertainties hinder drawing definitive conclusions. These observations may be related to the presence of the onset of deconfinement~\cite{Gazdzicki:2003bb, Gorenstein:2003hk} or a critical point~\cite{Sarkar:2025xpv}, but additional evidence is necessary. Further studies in Ar+Sc collisions require either increasing data statistics or finding more suitable quantities that can be applied to wider centrality bins, where volume fluctuations cannot be neglected~\cite{Sangaline:2015bma}.

\section{Direct measurement of open charm at SPS}
\label{sec:opench}

\begin{figure}[h]
\centering
\sidecaption
\includegraphics[width=7.25cm,clip]{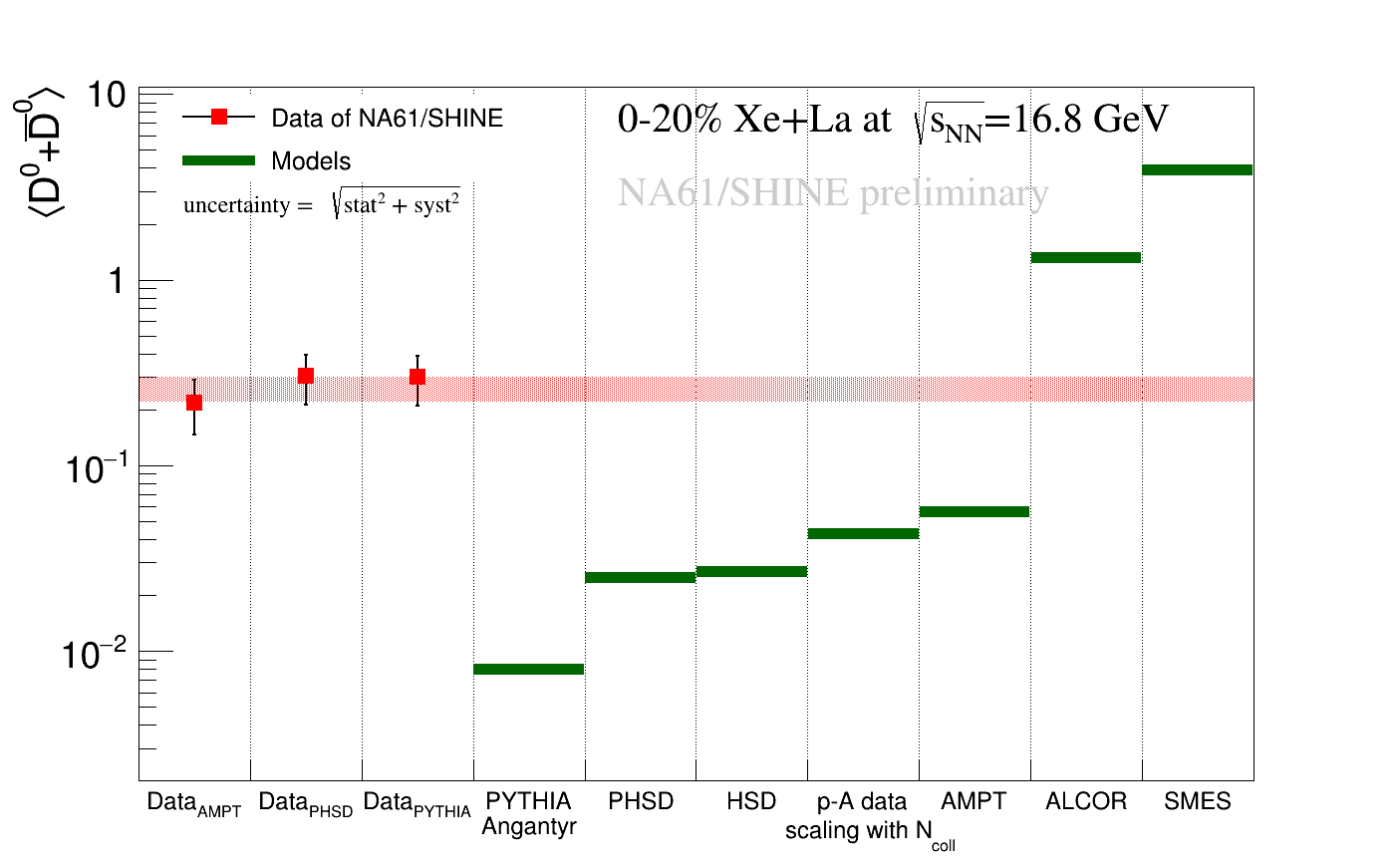}
\caption{\NASixtyOne preliminary results on open charm production in 0--20\% central Xe+La collisions at \\ \snn = 16.8~\GeV. The raw yields were corrected using AMPT, PHSD, and Pythia/Angantyr models. Vertical bars denote total uncertainties. See Ref.~\cite{Merzlaya:2024cbt} for numerical values (at $4\pi$ and mid-rapidity) and references to models. }
\label{fig:open_charm}
\end{figure}

Last year, \NASixtyOne reported the first-ever direct measurement of open charm production in nucleus-nucleus collisions at SPS. This was made possible thanks to a prototype version of the vertex detector for measuring short-lived particles. The raw yields in 0--20\% central Xe+La collisions at \snn = 16.8~\GeV were corrected using AMPT, PHSD, and Pythia/Angantyr models. The resulting mean multiplicities of open charm mesons are presented in Fig.~\ref{fig:open_charm} together with model predictions. The precision of the data is sufficient to discriminate between models. Microscopic models tend to underpredict the yields, whereas statistical models overpredict them. The analysis of high-statistics Pb+Pb data at 16.8~\GeV, collected since 2022 using the upgraded \NASixtyOne detector, is ongoing.

\section{Evidence of isospin-symmetry violation in kaon production}
\label{sec:isospin}

\begin{figure}[h]
\centering
\vspace{0.2cm}
\includegraphics[width=10cm,clip]{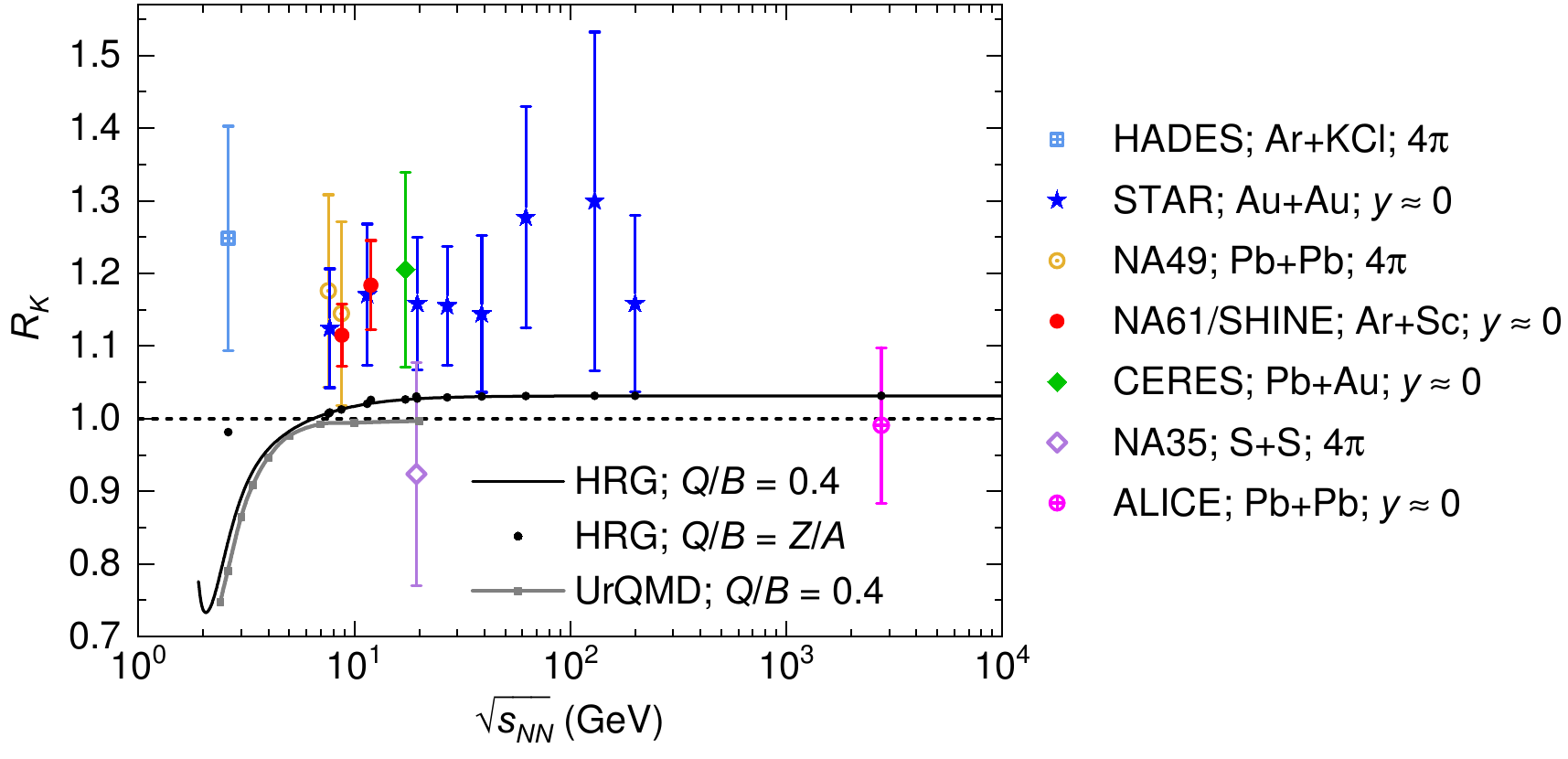}
\vspace{-0.3cm}
\caption{Ratio of charged to neutral $K$ meson yields in nucleus-nucleus collisions as a function of collision energy. The black line shows the Hadron Resonance Gas (HRG) model predictions for $Q/B=0.4$. The black dots indicate the HRG predictions for $Q/B$ values corresponding to the ones in the experiments. For different nuclei, $Q/B$ corresponds to the electric charge over the baryon number of the whole system. The gray squares show UrQMD predictions. Vertical bars denote total uncertainties. Figure taken from Ref.~\cite{NA61SHINE:2023azp} with new preliminary Ar+Sc 8.8~\GeV point added~\cite{QM2025_YB}. }
\label{fig:Rk_energy}
\end{figure}

\NASixtyOne has recently announced the observation of an unexpected excess of charged over neutral kaons produced in Ar+Sc collisions at \snn = 11.9~\GeV, which has been interpreted as evidence of a significant isospin-symmetry violation~\cite{NA61SHINE:2023azp}.
In the case of exact isospin symmetry and for collisions involving nuclei with equal numbers of protons and neutrons ($Z$ = $N$), the kaon multiplicities follow relations $\langle K^{+} \rangle (u \bar{s}) = \langle K^{0} \rangle (d \bar{s})$ and $\langle K^{-} \rangle (\bar{u} s) = \langle \overline{K}^{\,0} \rangle (\bar{d} s)$.  
Ignoring the small effects of \textit{CP} (charge-parity) violation, the average yields of short- and long-lived neutral kaons satisfy $\langle \Ks \rangle=\frac{1}{2} \langle K^{0} \rangle +\frac{1}{2} \langle \overline{K}^{\,0} \rangle = \langle \Kl \rangle$. 
Accordingly, the ratio
\begin{equation}
R_K\equiv\frac{\langle K^{+} \rangle  + \langle K^{-} \rangle}{\langle K^0 \rangle  + \langle \overline{K}^{\,0} \rangle} = \frac{\langle K^+ \rangle  + \langle K^- \rangle}{2 \langle \Ks \rangle }
\end{equation}
is expected to equal unity if isospin symmetry holds.
Nevertheless, \NASixtyOne data for the 10\% most central Ar+Sc collisions revealed a significant excess of charged kaons compared to neutral ones: approximately 18\% at $\sqrt{s_\mathrm{NN}}$ = 11.9~\GeV (published result~\cite{NA61SHINE:2023azp}) and about 12\% at$\sqrt{s_\mathrm{NN}}$ = 8.8~\GeV (preliminary~\cite{QM2025_YB}). Similar charged-kaon excesses had been observed in earlier nucleus-nucleus collision experiments (see Fig.~\ref{fig:Rk_energy}) but remain unaccounted for by existing theoretical models. The observed deviation corresponds to a violation of isospin symmetry at the level of 4.7$\sigma$ (rising to 5.3$\sigma$ when including the 8.8~\GeV data) beyond the effects expected from known isospin-breaking mechanisms (the significance is obtained based on the comparison with predictions of the Hadron Resonance Gas model, see Ref.~\cite{NA61SHINE:2023azp} for details). The analysis of Ar+Sc collisions at \snn = 16.8~\GeV is in progress.

\section{Future plans}
\label{sec:future}

As part of its strong interaction program, \NASixtyOne will continue recording Pb+Pb data at 16.8~\GeV until 2026, with the goal of studying open charm production. A total of 500 million minimum bias events are planned to be recorded in the years 2022--2026.
After the CERN Long Shutdown~3, collisions of Mg+Mg, O+O, and B+B are planned at \snn = 5.1, 7.6, and 16.8 \GeV. A pilot run with O+O at 16.8 GeV was recorded in 2025. These measurements aim to explore the system size dependence of the $K^{+}/\pi^{+}$ ratio and $T$,  and study isospin violation in isospin-symmetric ($Z$ = $N$) $^{16}_8$O+$^{16}_8$O and $^{24}_{12}$Mg+$^{24}_{12}$Mg collisions.
Finally, 24 million events for each of $\pi^+$+C and $\pi^-$+C interactions at beam momentum 158~\GeVc were recorded in 2024. The goal is to study isospin symmetry violation. \NASixtyOne reported $R_K > 1$ not only in central Ar+Sc collisions, but also in $\pi^-$+C interactions at beam momenta 158 and 350~\GeVc ~\cite{NA61SHINE:2022tiz}. The $\pi^-$+C interaction is not an isospin-symmetric system, but it will be interesting to check whether $\pi^+$+C will be the mirror image of $\pi^-$+C with respect to $R_K$ = 1, as expected from isospin symmetry.

\vspace{0.3cm}
\noindent
\small {\textbf{Acknowledgements:} This work was partially supported by the Polish Ministry of Science and Higher Education (grant WUT ID-UB)}.




\bibliography{references}

%
%

%
%

\end{document}